\documentclass[11pt,a4paper]{article}
\pdfoutput=1
\usepackage{jcappub}

\usepackage{appendix}
\usepackage{array}
\newcolumntype{x}[1]{>{\centering\arraybackslash}p{#1}}

\usepackage{epsfig}
\usepackage{graphicx}
\usepackage{dcolumn}
\usepackage{amsmath}
\usepackage{enumerate}
\def\lsim{\mathrel{\hbox{\rlap{\hbox{\lower4pt\hbox{$\sim$}}}\hbox{$<$}}}}

\makeatletter

\newcommand{\Rmnum}[1]{\expandafter\@slowromancap\romannumeral #1@}
\makeatother

\title{Effects of kination and scalar-tensor cosmologies on sterile neutrinos}

\author[a]{Thomas Rehagen}
\author[a]{and Graciela B. Gelmini}

\affiliation[a]{Department of Physics and Astronomy, UCLA,\\
475 Portola Plaza, Los Angeles, CA 90095, USA}

\emailAdd{trehagen@physics.ucla.edu}
\emailAdd{gelmini@physics.ucla.edu}


\abstract{We study the effects of kination and scalar-tensor pre-Big Bang Nucleosynthesis cosmologies on the non-resonant production of sterile neutrinos.  We show that if the peak of the production rate of sterile neutrinos occurs during a non-standard cosmological phase, the relic number density of sterile neutrinos could be reduced with respect to the number expected in the standard cosmology. Consequently, current bounds on active-sterile neutrino mixing derived from the relic energy density of sterile neutrinos could be greatly relaxed.  In particular, we show that the sterile neutrinos which could explain the anomalies found in short-baseline neutrino experiments are compatible with recent joint Planck upper limits on their contribution to the energy density of the Universe in a scalar-tensor or a low-reheating temperature pre-Big Bang Nucleosynthesis cosmology.}

\begin{document}

\maketitle

\section{Introduction}

In the Standard Model of particle physics, there are three active neutrinos, $\nu_\alpha$, coupled to the W and Z weak gauge bosons through gauge couplings.  Extensions of the Standard Model can include one or more sterile neutrinos, $\nu_s$, that are not directly coupled to the W and Z bosons, but are produced in the early Universe via mixing with the active neutrinos.  Barbieri and Dolgov~\cite{Barbieri} and Dodelson and Widrow~\cite{Dodelson:1993je} provided respectively the first analytical calculation of the production probability and the first analytic solution for the relic number density of sterile neutrinos produced non-resonantly in the early Universe, under the assumption of a negligible primordial lepton asymmetry (later corrected by a factor of 2~\cite{oai:arXiv.org:hep-ph/0202122, Abazajian:2001nj}).  If there is a large lepton asymmetry, sterile neutrinos can be produced resonantly \cite{Abazajian:2001nj,Shi:1998km} with a non-thermal spectrum which favors low energies. 

One or two sterile neutrinos with mass of $\mathcal{O}(1\,{\rm eV})$ have been proposed as a way to explain several anomalies observed in short-baseline neutrino experiments, which aim a beam of neutrinos at a nearby ($\lesssim 1\,{\rm km}$ away) detector.  LSND \cite{Aguilar:2001ty} and Mini-BooNE \cite{AguilarArevalo:2010wv, Aguilar-Arevalo:2013pmq} observed an excess of electron (anti-)neutrinos above the expected level. This excess could be explained by the oscillation of muon (anti-)neutrinos into one or two sterile (anti-)neutrinos, and the subsequent oscillation of these sterile (anti-) neutrinos into electron (anti-)neutrinos.  In addition to the LSND and Mini-BooNE results, radioactive source experiments meant to test the Gallium solar neutrino experiments SAGE and GALLEX have observed an event rate that is lower than expected.  This effect can be explained by electron neutrinos oscillating into sterile neutrinos \cite{Giunti:2010zu,Acero:2007su}.  A recent re-evaluation of the neutrino flux emitted by nuclear reactors \cite{Mueller:2011nm, Huber:2011wv} has led to increased predicted fluxes compared to previous calculations \cite{Schreckenbach:1985ep,Hahn:1989zr,VonFeilitzsch:1982jw,Vogel:1980bk}.  Previous reactor experiments have not observed this flux, a result which can be explained by assuming $\bar{\nu_e}$ disappearance due to oscillations with a sterile neutrino. The sterile neutrinos that could explain these anomalies have a best fit active-sterile mixing angle ${\rm sin}^22\theta \simeq 0.08$ (and ranging from $0.03$ to $0.4$ at the $95\%$ confidence level)~\cite{Kopp:2013vaa}.  From now on we will refer to sterile neutrinos with mass $m_{\nu_s}=1\,{\rm eV}$ and mixing angle ${\rm sin}^2(2\theta)=0.08$ as the ``LSND $\nu_s.$''

Recently, the Planck satellite experiment has provided joint constraints on the effective number of relativistic degrees of freedom at recombination, $N_{eff}$, and the effective mass of a sterile neutrino, $m_{\nu_s}^{eff}$, in different models for active and sterile neutrinos.  $N_{eff}$ is defined as
\begin{equation}
N_{eff}=3.046+\frac{\rho_{\nu_s}}{\rho_{\nu_\alpha}},
\label{neff}
\end{equation}
where $3.046$ is the expected value of $N_{eff}$ in the absence of sterile neutrinos \cite{Mangano:2005cc}.  $\rho_{\nu_s}/\rho_{\nu_\alpha}$ is the ratio of the energy density of relativistic sterile neutrinos, $\rho_{\nu_s}=2\int d^3p/(2\pi)^3E_{\nu_s}f_{\nu_s}$ for each $\nu_s$ species, and the energy density of a neutrino in thermal equilibrium is $\rho_{\nu_\alpha}$.  The effective sterile neutrino mass is defined for non-relativistic sterile neutrinos as 
\begin{equation}
m_{\nu_s}^{eff} = \frac{\sum  n_{\nu_s}m_{\nu_s}}{n_{\nu_\alpha}},
\label{meff}
\end{equation}
where the sum is over sterile neutrino species, and $n_{\nu_\alpha}$ is the number density of an active neutrino.  For relativistic sterile neutrinos in thermal equilibrium with the active neutrinos, each $\nu_s$ species would have the same relic density as one of the active neutrinos, thus $\rho_{\nu_s}/\rho_{\nu_\alpha}$ would in fact be the number of sterile neutrino species.  For a non-relativistic neutrino in thermal equilibrium before decoupling, its relic number density would be the same as the number density of an active neutrino, $n_{\nu_\alpha}$, and its relic energy density would be $\rho_{\nu_s}=m_{\nu_s}n_{\nu_\alpha}$.  Thus $m_{eff}$ would be the sum of the masses of all the sterile neutrino species.  

Assuming the existence of two massless active neutrinos and a massive active neutrino with mass $m\simeq 0.06\,{\rm eV}$ fixed by the atmospheric mass splitting, and a massive sterile neutrino with either a thermal distribution or a distribution proportional to that of the active neutrinos, and combining Planck, WMAP, Baryon Acoustic Oscillation, and high multipole CMB data, Planck finds the $95\%$ limits \cite{Ade:2013zuv}
\begin{equation}
N_{eff}<3.80 
\label{nval}
\end{equation}
and for sterile neutrinos with mass $m_{\nu_s}<10\,{\rm eV}$
\begin{equation}
m_{\nu_s}^{eff}<0.42\,{\rm eV}.
\label{mval}
\end{equation}
The conditions of these bounds apply to the sterile neutrino distributions we consider in this paper.  The measurement of $N_{eff}$ provides a bound on the mixing angle of sterile neutrinos that are relativistic at the time of recombination, i.e. with mass $m_{\nu_s} \lesssim 1\,{\rm eV}$, while sterile neutrinos with higher masses up to $10\,{\rm eV}$ are constrained by the measurement of $m_{\nu_s}^{eff}$.  Refs.~\cite{Mirizzi:2013kva} and \cite{Valentino:2013wha} have applied these bounds to a 3 active + 1 sterile neutrino model, and find the sterile neutrino needed to explain short-baseline neutrino experiment anomalies is highly disfavored.  Indeed, Ref.~\cite{Mirizzi:2013kva} finds that this sterile neutrino is excluded at more than $4\sigma$ by the Planck results.

One way to relax these bounds is to have a reduced number of sterile neutrinos produced in the early Universe.  Many possible scenarios have been considered, including models with a  non-standard cosmology, such as a low reheating temperature\cite{Gelmini:2004ah, Gelmini:2008fq}, a constant $w\neq 1$ dark energy equation of state \cite{Hannestad:2005gj,Hamann:2011ge,Giusarma:2011zq}, a time-varying dark energy component \cite{Giusarma:2011zq}, and interactions between the dark matter and dark energy sectors \cite{Giusarma:2011zq,Kristiansen:2009yx,Gavela:2009cy}, or non-standard particle physics such as models with a large neutrino asymmetry \cite{Foot:1995bm,Chu:2006ua,Saviano:2013ktj}, and interactions between sterile neutrinos and non-Standard Model particles \cite{Bento:2001xi, Dasgupta:2013zpn,Hannestad:2013ana}.  

Heavier sterile neutrinos with mass $m_{\nu_s}\gtrsim {\rm keV}$ could account for the whole of the dark matter.  In addition, if such sterile neutrinos carry away a sizable fraction of the energy emitted in a supernova explosion, asymmetric emission of the sterile neutrinos due to the presence of a strong magnetic field could explain the very large velocities of pulsars \cite{Fuller:2003gy}.  The mass and mixing of sterile neutrinos that make up all of the dark matter is subject to constraints from X-ray measurements, phase space density arguments, and analysis of the Lyman-$\alpha$ forest data.  If the keV mass sterile neutrinos are produced non-resonantly, they are strongly disfavored as the sole dark matter component \cite{Boyarsky:2008ju}.  If these sterile neutrinos are not the primary dark matter component, however, these bounds are weakened \cite{Palazzo:2007gz}, and a new region of parameter space could be opened, as long as the relic energy density of the keV mass sterile neutrinos does not exceed the energy density of dark matter.

This paper considers the effects of kination and scalar-tensor pre-Big Bang Nucleosynthesis (BBN) cosmologies on sterile neutrinos.  In both of these cosmologies, the Hubble expansion rate is greater than in the standard cosmology, and so the production of sterile neutrinos in the early Universe is suppressed.  For comparison we also include the results found in Ref.~\cite{Gelmini:2004ah} for sterile neutrinos evolving in a low reheating temperature cosmology.

\section{Non-standard pre-BBN cosmologies}

In the standard cosmology, the early Universe is radiation dominated, and the expansion rate of the Universe is given by $H_{STD}=\pi T^2/M_P\sqrt{g_\star/90}$, where $T$ is the temperature of the radiation bath, $M_P$ is the reduced Planck mass, and $g_\star$ is the number of relativistic degrees of freedom.   For temperatures higher thant the QCD phase transition ($\sim 200\,{\rm MeV}$), $g_\star$ is approximately constant, with a value of $g_\star=90$, while for temperatures between the QCD phase transition and the temperature at the moment electrons and positrons become non-relativistic ($T \simeq \,{\rm MeV}$), $g_\star=10.75$.  In order for BBN and the subsequent history of the Universe to develop as usual, the Universe must be radiation dominated for $T<4 \, {\rm MeV}$ at the $95\%$ confidence level \cite{Hannestad:2004px} (see also \cite{DeBernardis:2008zz, Kawasaki:2000en, Kawasaki:1999na}), but different cosmologies are allowed at higher temperatures. 

In the kination model \cite{Spokoiny:1993kt,Joyce:1996cp,Salati:2002md,Profumo:2003hq,Pallis:2005hm}, there is a period in which the kinetic energy of a scalar field $\phi$ dominates over its potential energy and all other contributions to the total energy density $\rho_{total}$, so $\rho_{total}\simeq \rho_\phi \simeq \dot{\phi}^2/2 \sim a^{-6}$, where $a$ is the scale factor of the Universe.  The entropy of matter plus radiation is conserved, so $a \sim T^{-1}$, thus the expansion rate of the universe is $H_{K}\sim \sqrt{\rho_{total}}\sim T^3$.  The contribution of the $\phi$ kinetic energy to the total energy density is often quantified through the ratio of $\phi$-to-photon energy density, $\eta_\phi=\rho_\phi/\rho_\gamma$ at $T=1\,{\rm MeV}$, so that at higher temperatures, $H_{K}\simeq \sqrt{\eta_\phi}(T/1\,{\rm MeV})H_{STD}$, where $H_{STD}$ is the standard expansion rate.  We can find $\eta_\phi$ by assuming that there is a rapid transition between the kination phase and the radiation dominated phase that occurs at the transition temperature $T_{tr}$.  In this case, $H_{K}(T_{tr})=H_{STD}(T_{tr})$, and we find that during the kination phase, the expansion rate of the Universe is given by $H_K=\pi T^3/(M_P T_{tr})\sqrt{g_{\star}/90}$.

Scalar-tensor theories of gravity \cite{Santiago:1998ae,Catena:2004ba} have a scalar field coupled through the metric tensor to the matter fields.  This scalar field changes the expansion rate of the Universe before the thermal bath has the transition temperature $T_{tr}$, after which the theory is indistinguishible from General Relativity.  In Ref.~\cite{Catena:2004ba}, the expansion rate of the Universe before the transition, $H_{ST}$, is proven to be enhanced with respect to $H_{STD}$ by a factor $f_\phi\simeq 2.19\times 10^{14}(T_0/T)^{0.82}$ (where $T_0$ is the present temperature of the Universe) at temperatures greater than $T_{tr}$.  At $T_{tr}$, $f_\phi$ drops sharply to values close to 1.

\begin{figure}
\begin{center}
\includegraphics[width=10cm]{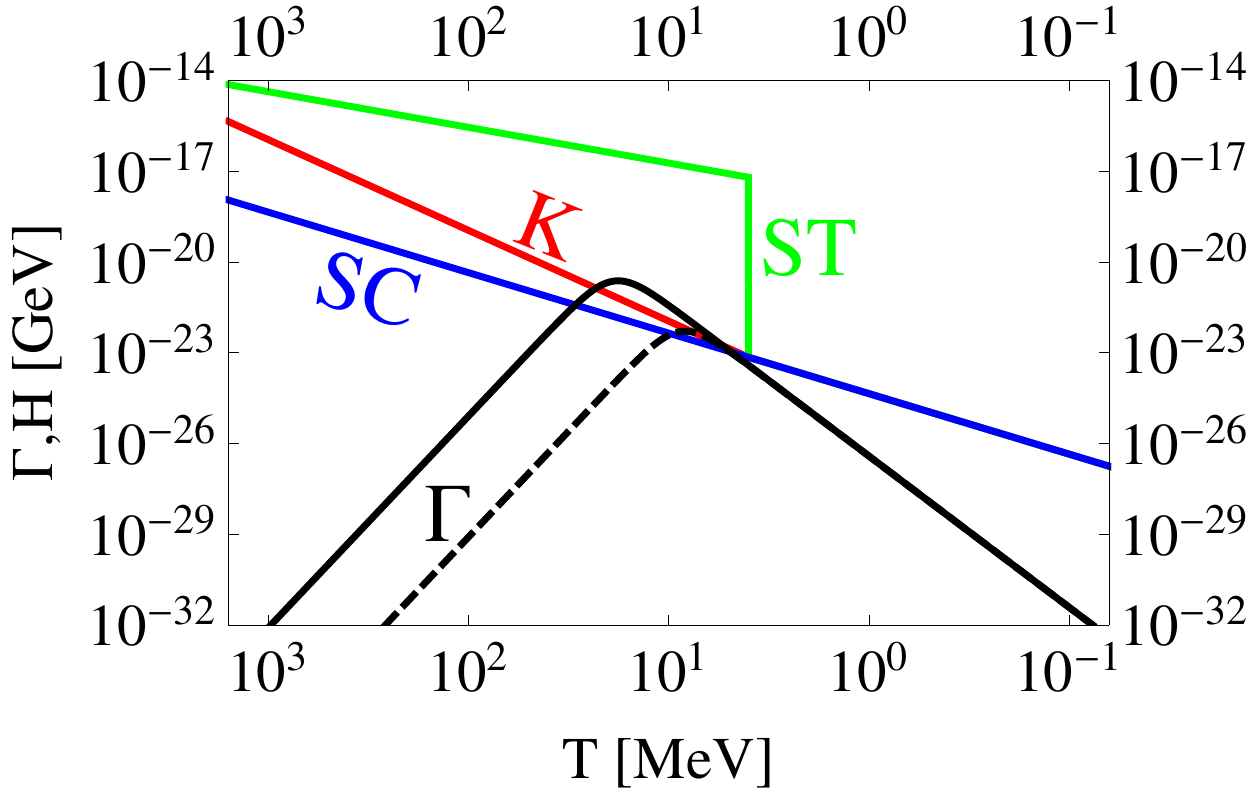}
\caption{Expansion rate of the Universe $H$ as a function of temperature for the standard cosmology, STD (blue), and for the kination, K (red), and scalar-tensor, ST (green), cosmologies with transition temperature $T_{tr}=4\,{\rm MeV}$ to the standard cosmology.  Also plotted is the rate of sterile neutrino interactions $\Gamma$ in Eq.~\ref{rate} for sterile neutrinos mixing with electron neutrinos with mixing angle ${\rm sin}^2(2\theta)=0.1$ and mass $m_{\nu_s}=1\,{\rm eV}$ and $0.1\,{\rm eV}$ (solid black and dashed black lines, respectively).}
\label{fig:Hubble}
\end{center}
\end{figure}

Fig.~\ref{fig:Hubble} shows the expansion rate of the Universe $H$ as a function of temperature for the standard cosmology, $H_{STD}$ (blue), and for the kination, $H_{K}$ (red), and scalar-tensor, $H_{ST}$ (green), cosmologies with transition temperature $T_{tr}=4\,{\rm MeV}$ to the standard cosmology.

In low reheating temperature models, a scalar field $\phi$ oscillating around its true minimum while decaying dominates the energy density of the Universe just before Big Bang Nucleosynthesis.  The decay of $\phi$ and the thermalization of the decay products reheats the thermal bath to a ``reheating temperature," $T_R$.  As we mentioned above, the reheating temperature can be as low as  $4\,{\rm MeV}$.  After the decay of $\phi$, radiation dominates the energy density of the Universe.  This case, which we include only for comparison, was studied in Refs.~\cite{Gelmini:2004ah} and \cite{Gelmini:2008fq}.

\section{Relic densities}
\label{3}

In this section we calculate the relic number and energy densitiy of only one species of sterile neutrinos, $\nu_s$, produced in the kination and scalar-tensor pre-BBN cosmologies.  We assume that there is a negligible primordial lepton asymmetry, that the active neutrinos have the usual thermal equilibrium distribution $f_{\nu_\alpha}=({\rm exp}\,E/T+1)^{-1}$, and that the sterile neutrinos are relativistic when produced.  We assume also that the sterile neutrino species interacts dominantly with only one active neutrino $\nu_\alpha$, and we use $\alpha$ to denote its charged lepton partner.  Under these assumptions, the Boltzmann equation for the production of sterile neutrinos via oscillations with active neutrinos is given in Ref.~\cite{Abazajian:2001nj}:
\begin{equation}
-HT \left(\frac{\partial f_{\nu_s}(E,T)}{\partial T} \right)_{E/T} \simeq \Gamma_s (E,T) \left[ f_{\nu_\alpha} (E,T)-f_{\nu_s}(E,T) \right],
\label{boltzmann}
\end{equation}
where the rate of sterile neutrino interactions is 
\begin{equation}
\Gamma_s(E,T) \simeq \frac{1}{4}{\rm sin}^2(2\theta_m)d_\alpha G_F^2ET^4,
\label{rate}
\end{equation}
assuming that the time between interactions is always much greater than the oscillation time, $t_{osc}=m^2_{\nu_s}/2E$, so that ${\rm sin}^2(t/t_{osc})$ averages to $1/2$.  The subscript $E/T$ on the partial derivative of the phase space distribution with respect to temperature indicates that the derivative is to be taken while holding $E/T$ constant.  Here $G_F$ is the Fermi constant, $d_\alpha=1.13$ for sterile neutrinos mixing with electron neutrinos ($\nu_s \leftrightarrow \nu_\alpha =\nu_e$) and $d_\alpha = 0.79$ for $\nu_\alpha = \nu_{\mu,\tau}$, and $\theta_m$ is the angle that parametrizes the magnitude of the mixing between the sterile and active neutrinos in the presence of matter effects given by \cite{Abazajian:2001nj}
\begin{equation}
{\rm sin}^2(2\theta_m)=\frac{{\rm sin}^2(2\theta)}{{\rm sin}^2(2\theta)+\left[ {\rm cos}(2\theta) - 2E (V_T)/m_{\nu_s}^2\right]^2}.
\label{sin2th}
\end{equation}
In this equation $\theta$ is the mixing angle in vacuum, $m_{\nu_s}$ is the mass of the sterile neutrino, and $V_T$ is the thermal potential.  If the active neutrino $\nu_\alpha$ is in thermal equilibrium, the thermal potential has the form $V_T=-BE\left(T/{\rm GeV} \right)^4,$ where $B=10.88 \times 10^{-9}$ when $\alpha$ is in thermal equilibrium and $B=3.02 \times 10^{-9}$ after $\alpha$ decouples from the thermal bath.  For $\alpha = e$, the electrons are always coupled to the thermal bath in the era we consider, while for $\alpha = \mu$ or $\tau $, the decoupling occurs at $T\simeq 20\, {\rm MeV}$ and $T\simeq 180\, {\rm MeV}$, respectively.  Assuming that the mixing angle in vacuum is small, we adopt in our calculations ${\rm sin}^2(2\theta_m)\simeq{\rm sin}^2(2\theta)/\left[ 1 + 2BE^2(T/{\rm GeV})^4 /m_{\nu_s}^2\right]^2.$  For vacuum mixing angles as high as ${\rm sin}^2(2\theta)=0.1$, this approximation leads to an error of at most 10 percent in the denominator of Eq.~\ref{sin2th}.  Note that as $T$ increases, the thermal potential becomes important when $1\simeq 2BE^2(T/{\rm GeV})^4 /m_{\nu_s}^2$.  Taking $E \simeq T$, we find that thermal effects are important for $ T \gtrsim 19\, {\rm GeV} (m_{\nu_s}/{\rm keV})^{1/3}$.

In Fig.~\ref{fig:Hubble} we show the rate of sterile neutrino interactions $\Gamma_s(E,T)$ for sterile neutrinos mixing with electron neutrinos with mixing angle ${\rm sin}^2(2\theta)=0.1$ and mass $m_{\nu_s}=1\,{\rm eV}$ and $0.1\,{\rm eV}$ (solid black and dashed black lines, respectively), the parameters relevant to explain the anomalies in short-baseline neutrino experients.  We see that in the standard cosmology, the production of these sterile neutrinos is in equilibrium only close to the maximum of the interaction rate.

Since the expansion rate of the universe $H$ is greater in the kination and scalar-tensor cosmologies than the expansion rate in the standard cosmology prior to BBN (see Fig.~\ref{fig:Hubble}), Eq.~\ref{boltzmann} makes clear that the production rate of sterile neutrinos in the non-standard cosmologies is lower than in the standard cosmology.  We therefore expect the relic number density of sterile neutrinos in the kination or scalar-tensor models to be smaller than the relic density in the standard cosmology.

\begin{figure}
\begin{center}
\includegraphics[width=10cm]{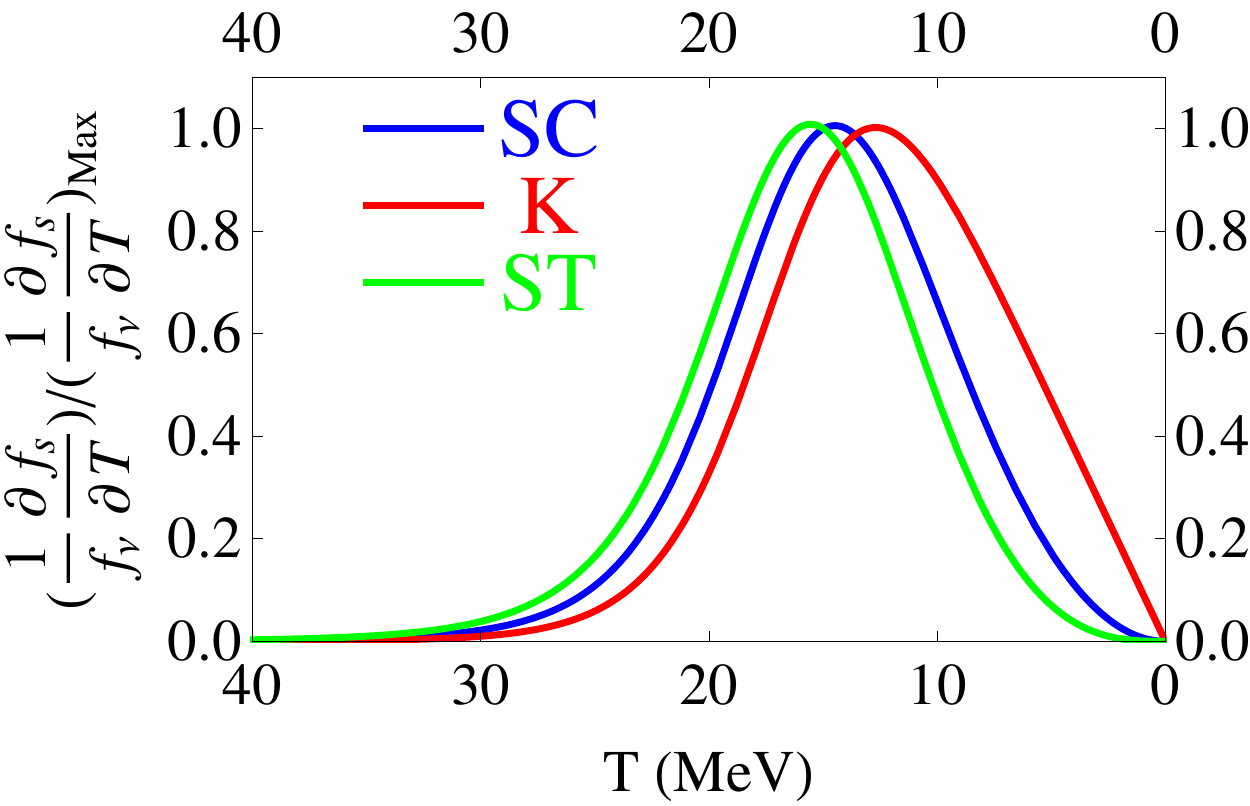}
\caption{$(\partial f_{\nu_s}/\partial T)_{E/T}$ when $f_{\nu_s}\ll f_{\nu_\alpha}$ for sterile neutrinos of mass $m_{\nu_s} = 1\,{\rm eV}$ as a function of temperature in the standard cosmology (blue line), kination model (red line), and scalar-tensor model (green line).  The peak rate of production has been normalized to 1 in each case.  Here we considered sterile neutrino production via $\nu_s \leftrightarrow \nu_e$ oscillations.  We approximate $g_\star = 10.75$.}
\label{fig:1}
\end{center}
\end{figure}

As shown in Fig.~\ref{fig:1}, the production rate $(\partial f_{\nu_s} / \partial T )_{E/T}$ has a sharp peak as a function of $T$.  The temperature $T_{max}$ at which $(\partial f_{\nu_s} / \partial T )_{E/T}$ peaks in the standard cosmology is $14.0\,{\rm MeV}(m_{\nu_s}/{\rm eV})^{1/3}$, while it is $13.0\,{\rm MeV}(m_{\nu_s}/{\rm eV})^{1/3}$ and $15.5\,{\rm MeV}(m_{\nu_s}/{\rm eV})^{1/3}$ for kination and scalar-tensor models, respectively, assuming that the peak occurs in the non-standard cosmological phase.  These values are for $B=10.88\times10^{-9}$, which assumes that the period of peak sterile neutrino production occurs when $\alpha$ is in thermal equilibrium.  If $\alpha$ is no longer in equilibrium during the peak production, then $T_{max}$ is increased by a factor of $(10.88/3.02)^{1/6}\simeq1.2$.  In order for the peak production rate to occur during the non-standard cosmological phase, i.e. $T_{max}\geq 4\,{\rm MeV}$, the sterile neutrino mass must be $m_{\nu_s}\gtrsim .01\,{\rm eV}$.

 Fig.~\ref{fig:1} shows $(\partial f_{\nu_s} / \partial T )_{E/T}$ when $f_{\nu_s}\ll f_{\nu_\alpha}$ for sterile neutrinos of mass $m_{\nu_s}=1\,{\rm eV}$ as a function of temperature in the standard cosmology (blue line), kination model (red line), and scalar-tensor model (green line).  The peak rate of production has been normalized to 1 in each case.  Fig.~\ref{fig:1} corresponds to sterile neutrinos produced via oscillations with electron neutrinos ($\nu_s \leftrightarrow \nu_e$).  In this case $g_\star = 10.75$, which is a good estimation when the peak production occurs between the QCD phase transition ($\simeq 200\,{\rm MeV}$) and the moment electrons and positrons become non-relativistic ($\simeq{\rm MeV}$).  If the peak production occurs before the QCD phase transition, we should take $g_\star=90$.  Using the expressions for $T_{max}$ mentioned above, and assuming for simplicity that the QCD phase transition occurs instantaneously at $T=200\,{\rm MeV}$, we see that the transition between $g_\star =90$ and $g_\star=10.75$ occurs at $m_{\nu_s}\simeq 3\,{\rm keV}$, $m_{\nu_s}\simeq 3.6\,{\rm keV}$, and $m_{\nu_s}\simeq 2.1\,{\rm keV}$ in the standard cosmology, kination cosmology, and scalar-tensor cosmology, respectively.

We find that as $T$ decreases in the standard cosmology $(\partial f_{\nu_s}/\partial T)_{E/T}$ increases as $T^{-10}$ for $T\gg T_{max}$ and decreases as $T^2$ for $T\ll T_{max}$.  Similarly, $(\partial f_{\nu_s}/\partial T)_{E/T}$ increases as $T^{-11}$ ($T^{-9.18}$) for $T\gg T_{max}$ and decreases as $T^1$ ($T^{2.82}$) for $T\ll T_{max}$ in the kination (scalar-tensor) model.  The quick drop-off from the peak production rate as $T$ decreases motivates the final approximation needed to solve Eq.~\ref{boltzmann}, i.e. to take $T_0\ll T_{max}$ in the standard cosmology and $T_{tr}\ll T_{max}$ in the non-standard cosmologies, since the majority of sterile neutrinos are produced near $T_{max}$.

Assuming that $f_{\nu_\alpha}\gg f_{\nu_s}$, we can rewrite Eq.~\ref{boltzmann} as
\begin{equation}
\left(\frac{\partial f_{\nu_s}(E,T)}{\partial T} \right)_{y} \simeq -\frac{1}{4}d_\alpha G_F^2 {\rm sin}^2(2\theta)\frac{y}{\left[1+2By^2T^6/m_{\nu_s}^2({\rm GeV})^4\right]^2} \frac{T^4}{H}f_{\nu_\alpha},
\label{explain1}
\end{equation}
where $y=E/T$ is held constant.  To find the phase space distribution in each cosmology, we must substitute the appropriate functional form of $H$ and integrate the right side of Eq.~\ref{explain1} over $T$ from $\infty$ to $T_0\simeq 0$.  Once the phase space distribution $f_{\nu_s}$ is obtained, we can compute the relic number density $n_{\nu_s}=2\int d^3p/(2\pi)^3f_{\nu_s}$ and relativistic energy density $\rho_{\nu_s}=2\int d^3p/(2\pi)^3pf_{\nu_s}$.  Using $f_{\nu_\alpha}\simeq {\rm e}^{-E/T}$, this gives for $T\ll T_{max}$

\begin{equation}
\frac{f_{\nu_s}^{STD}}{f_{\nu_\alpha}}=\frac{n_{\nu_s}^{STD}}{n_{\nu_\alpha}}=\frac{\rho_{\nu_s}^{STD}}{\rho_{\nu_\alpha}}\simeq 46\, d_\alpha {\rm sin}^2(2\theta)\left(\frac{90}{g_\star}\right)^{1/2}\left(\frac{10.88\times 10^{-9}}{B}\right)^{1/2}\frac{m_{\nu_s}}{\rm eV}
\label{nSTD}
\end{equation}
for the standard cosmology, where $n_{\nu_\alpha}=(2/\pi^2)T^3$ and $\rho_{\nu_\alpha}=(6/\pi^2)T^4$.



The sterile neutrino phase space distribution in the non-standard cosmologies can be found similarly. Since most of the sterile neutrino production occurs during the non-standard cosmology, we integrate Eq.~\ref{explain1} from $\infty$ to $T_{tr}\simeq 0$, and we find that for $T\ll T_{max}$
\begin{equation}
\frac{f_s^{K}}{f_{\alpha}}\simeq 15 \,d_\alpha {\rm sin}^2(2\theta)\left(\frac{90}{g_\star}\right)^{1/2}\left(\frac{10.88\times 10^{-9}}{B}\right)^{1/3}\left(\frac{T_{tr}}{4\,{\rm MeV}}\right)\left(\frac{m_{\nu_s}}{\rm eV}\right)^{2/3}\left(\frac{E}{T}\right)^{1/3}
\label{solKin}
\end{equation}
in the kination cosmology.  The relic number density is then
\begin{equation}
\frac{n_{\nu_s}^{K}}{n_{\nu_\alpha}}\simeq 21 \,d_\alpha {\rm sin}^2(2\theta)\left(\frac{90}{g_\star}\right)^{1/2}\left(\frac{10.88\times10^{-9}}{B}\right)^{1/3} \left(\frac{T_{tr}}{4\,{\rm MeV}}\right)\left(\frac{m_{\nu_s}}{\rm eV}\right)^{2/3},
\label{nKin}
\end{equation}
and the energy density of relativistic sterile neutrinos is
\begin{equation}
\frac{\rho_{\nu_s}^{K}}{\rho_{\nu_\alpha}}\simeq 24 \,d_\alpha {\rm sin}^2(2\theta)\left(\frac{90}{g_\star}\right)^{1/2}\left(\frac{10.88\times10^{-9}}{B}\right)^{1/3} \left(\frac{T_{tr}}{4\,{\rm MeV}}\right)\left(\frac{m_{\nu_s}}{\rm eV}\right)^{2/3}.
\label{nKin}
\end{equation}
In the scalar-tensor model instead,
\begin{equation}
\frac{f_s^{ST}}{f_\alpha}\simeq 1.5\times10^{-4}\, d_\alpha {\rm sin}^2(2\theta)  \left(\frac{90}{g_\star}\right)^{1/2}\left(\frac{10.88\times10^{-9}}{B}\right)^{3.82/6}\left(\frac{m_{\nu_s}}{\rm eV}\right)^{3.82/3}\left(\frac{E}{T}\right)^{-0.82/3},
\label{solST}
\end{equation}
\begin{equation}
\frac{n_{\nu_s}^{ST}}{n_{\nu_\alpha}}\simeq 1.2\times10^{-4}\, d_\alpha {\rm sin}^2(2\theta)\left(\frac{90}{g_\star}\right)^{1/2}\left(\frac{10.88\times10^{-9}}{B}\right)^{3.82/6}\left(\frac{m_{\nu_s}}{\rm eV}\right)^{3.82/3},
\label{nST}
\end{equation}
and
\begin{equation}
\frac{\rho_{\nu_s}^{ST}}{\rho_{\nu_\alpha}}\simeq 5.4\times10^{-5}\, d_\alpha {\rm sin}^2(2\theta)\left(\frac{90}{g_\star}\right)^{1/2}\left(\frac{10.88\times10^{-9}}{B}\right)^{3.82/6}\left(\frac{m_{\nu_s}}{\rm eV}\right)^{3.82/3}.
\label{nST}
\end{equation}

For comparison, we include the fraction of sterile and active neutrinos relic number densities in a low reheating temperature cosmology found in Eq.~2 of Ref.~\cite{Gelmini:2004ah}:
\begin{equation}
\frac{n_{\nu_s}^{LRT}}{n_{\nu_\alpha}}\simeq 10\,d_\alpha \,{\rm sin}^2(2\theta)\left(\frac{T_R}{5\,{\rm MeV}}\right)^3.
\end{equation}

\begin{figure}
\begin{center}
\includegraphics[width=10cm]{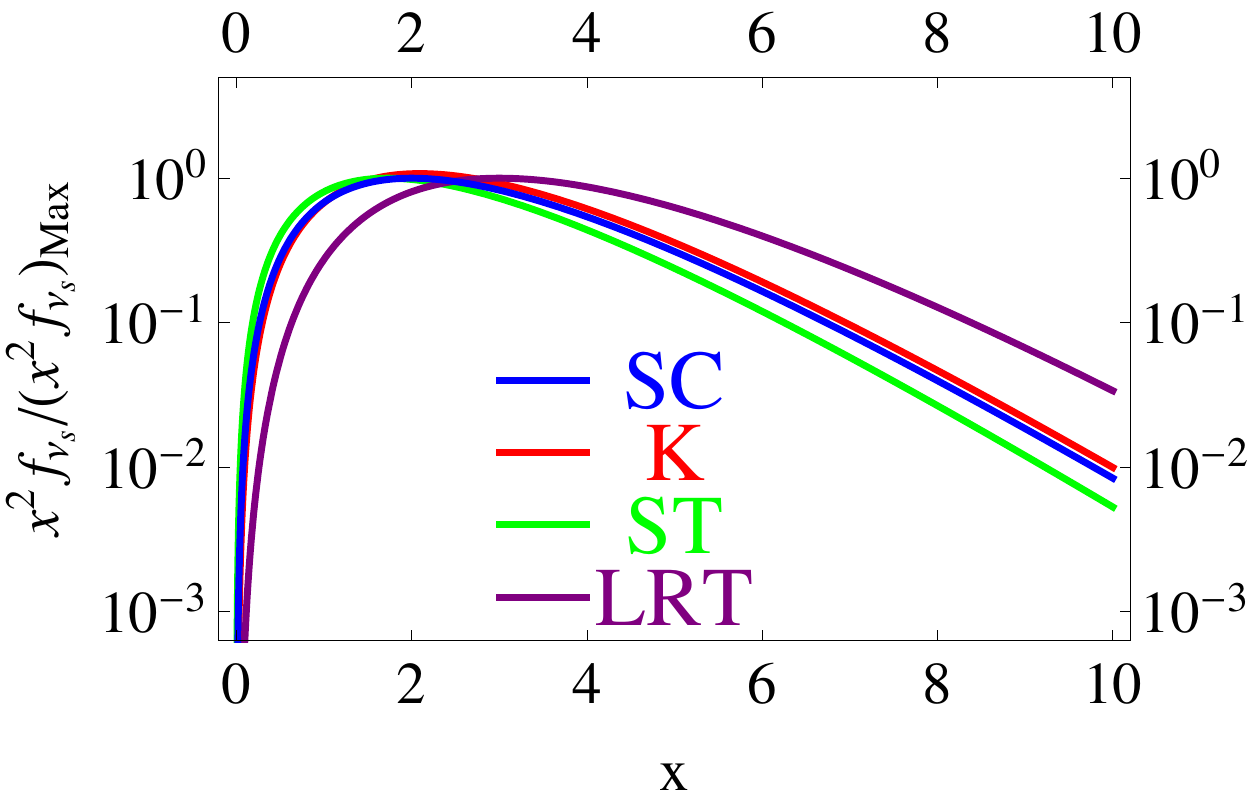}
\caption{$x^2 f_{\nu_s}(x)$, where $x=E/T$, for the LSND $\nu_s$ coupled to electron neutrinos in different pre-BBN cosmologies.  We have taken $T_{tr}=4\,{\rm MeV}.$  The peak of $x^2 f_{\nu_s}(x)$ has been normalized to 1 in each case.}
\label{fgraph}
\end{center}
\end{figure}

Eqs.~\ref{nSTD}, \ref{solKin}, and \ref{solST} are only valid if the condition $f_{\nu_s}\ll f_{\nu_\alpha}$ is satisfied.  These solutions can be extended to the case when this approximation begins to fail, which happens when the mixing angle becomes large enough.  In the case that $f_{\nu_s} \simeq f_{\nu_\alpha}$, the phase space density of sterile neutrinos is
\begin{equation}
\frac{f_{\nu_s}}{f_{\nu_\alpha}}\simeq 1-{\rm e}^{-S},
\label{extended}
\end{equation}
where $S=f_{\nu_s}/f_{\nu_\alpha}$ given in Eqs.~\ref{nSTD}, \ref{solKin}, and \ref{solST}.  This phase space distribution can then be integrated numerically to find the number density and energy density of sterile neutrinos.

Fig.~\ref{fgraph} shows $x^2 f_{\nu_s}(x)$, where $x=E/T$, for the LSND $\nu_s$ coupled to electron neutrinos for different pre-BBN cosmologies, Eqs.~\ref{nSTD}, \ref{solKin}, and \ref{solST} and Eq.~1 of Ref.~\cite{Gelmini:2004ah}.  We have taken $T_{tr}=4\,{\rm MeV}.$  The peak of $x^2 f_{\nu_s}(x)$ has been normalized to 1 in each case.  As can be seen in Fig.~\ref{fgraph}, the phase space distribution of sterile neutrinos evolving in these cosmologies is always close to a thermal equilibrium distribution.  Thus, the interactions between the sterile and active neutrinos should not significantly affect the distribution of the active neutrinos, which will retain a thermal spectrum.

\begin{figure}
\begin{center}
\includegraphics[width=10cm]{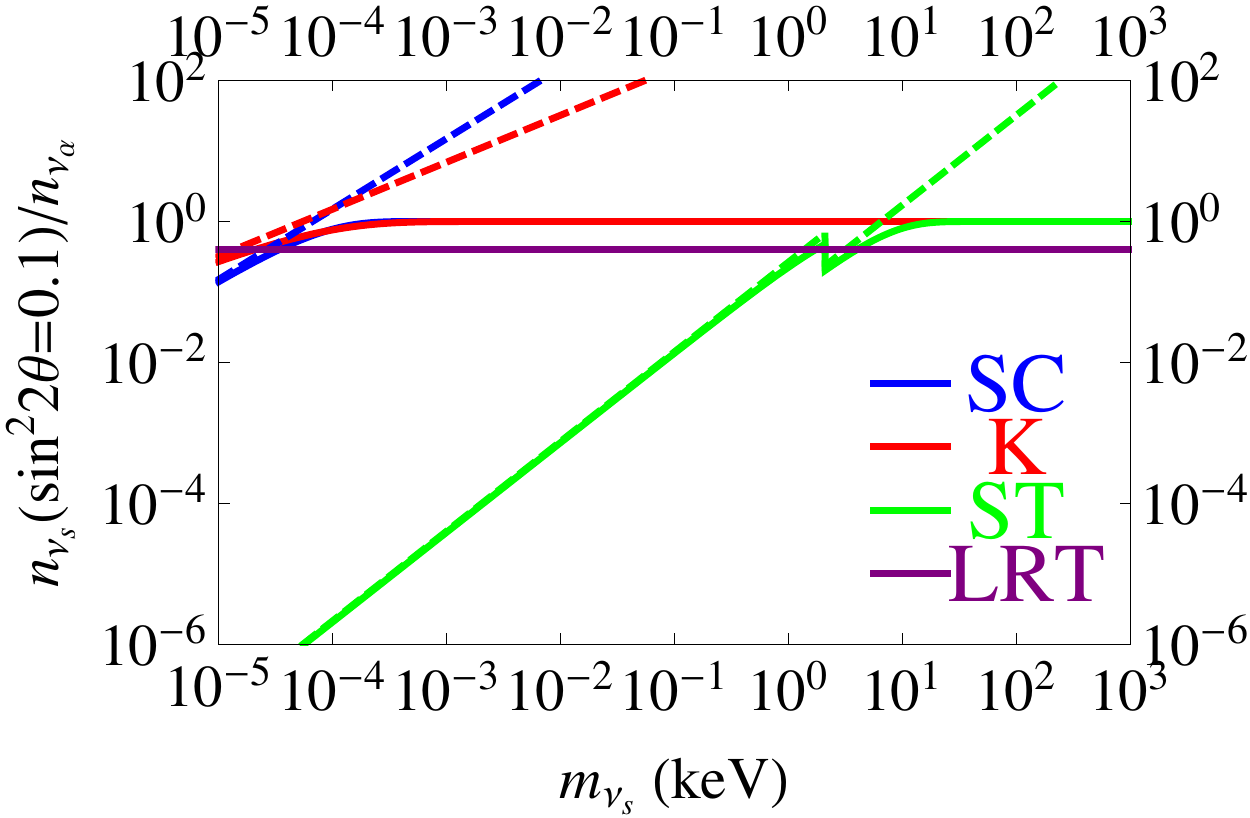}
\caption{$n_{\nu_s}/n_{\nu_\alpha}$ as a function of sterile neutrino mass $m_{\nu_s}$ mixing with active electron neutrinos with ${\rm sin}^22\theta=0.1$ for the standard cosmology (blue), the kination model with $T_{tr}=4\,{\rm MeV}$ (red), the scalar-tensor model (green), and the low reheating temperature model with $T_R=4\,{\rm MeV}$ (purple).   The solid lines show the number density found by integrating Eq.~\ref{extended}, while the dashed lines show the approximate analytic solutions valid when $f_{\nu_s}\ll f_{\nu_\alpha}$.  The kink in the lines at $m_{\nu_s}\simeq 3\,{\rm keV}$ is due to the approximation of an instantaneous change from $g_\star=90$ to $g_\star=10.75$ at the QCD phase transition.}
\label{fig:2}
\end{center}
\end{figure}

\begin{figure}
\begin{center}
\includegraphics[width=10cm]{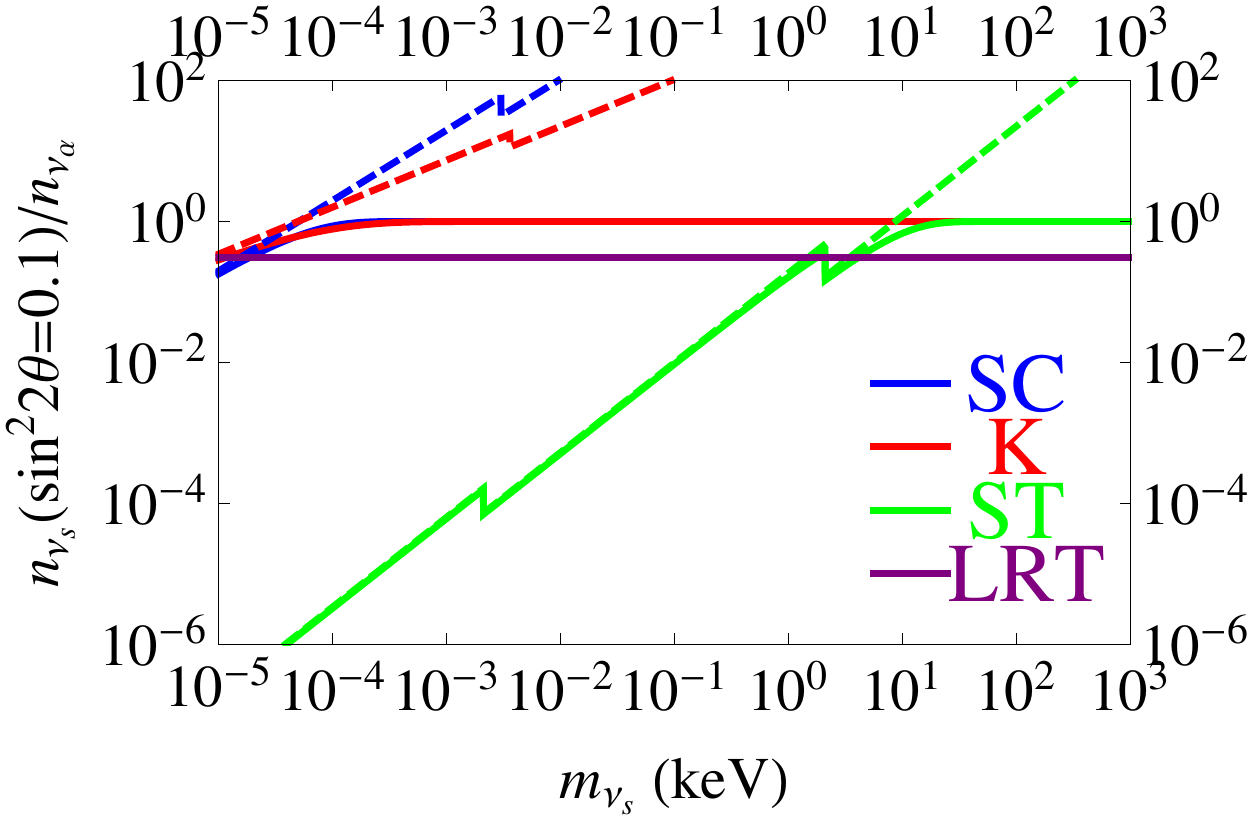}
\caption{Same as Fig.~\ref{fig:2} but for sterile neutrinos produced via oscillations with muon neutrinos ($\nu_s \leftrightarrow \nu_\mu$).  The new kink at $m_{\nu_s}\simeq 3 \times 10^{-3}\,{\rm keV}$ is due to the transition from $B=10.88\times 10^{-9}$ to $B=3.02\times 10^{-9}$ at $T\simeq 20 \, {\rm MeV}$.}
\label{fig:3}
\end{center}
\end{figure}

\begin{figure}
\begin{center}
\includegraphics[width=10cm]{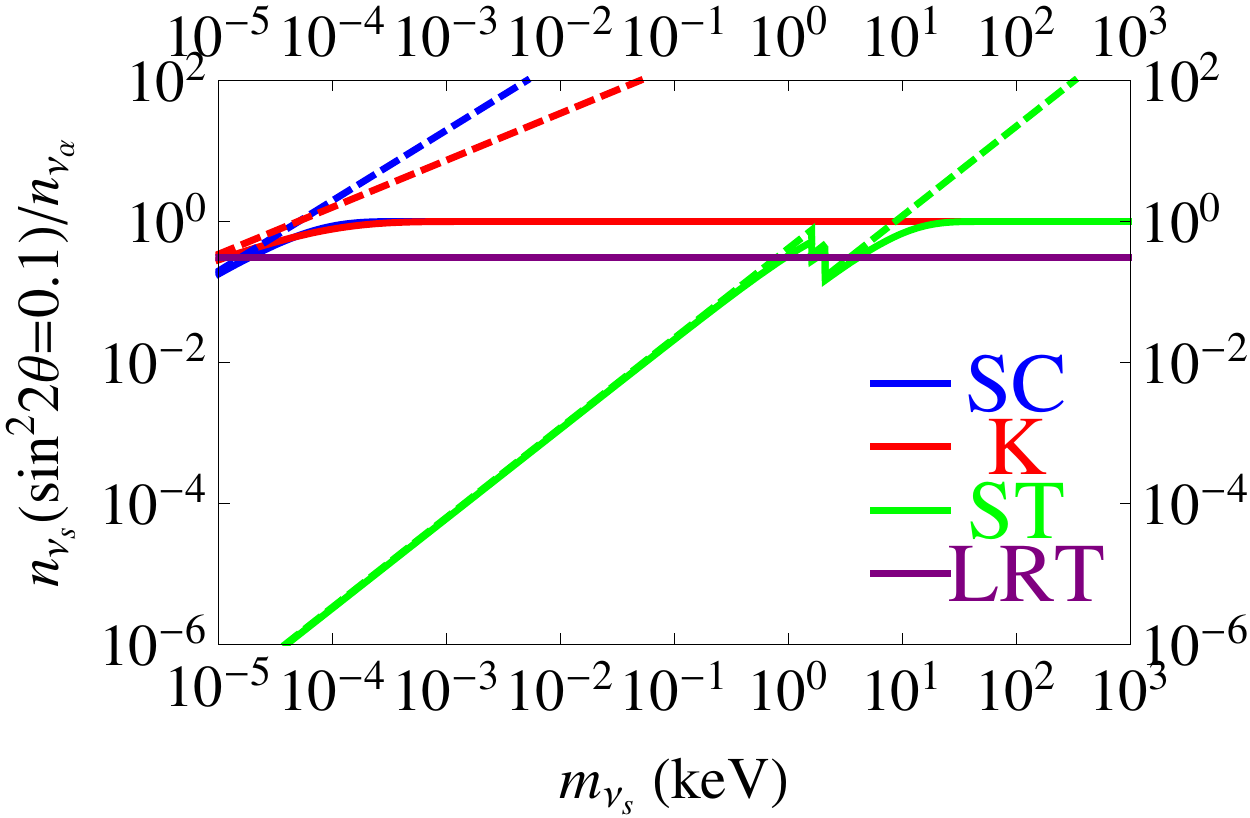}
\caption{Same as Fig.~\ref{fig:2} but for sterile neutrinos produced via oscillations with tau neutrinos ($\nu_s \leftrightarrow \nu_\tau$).  The new kink at $m_{\nu_s}\simeq 2 \,{\rm keV}$ is due to the transition from $B=10.88\times 10^{-9}$ to $B=3.02\times 10^{-9}$ at $T\simeq 180 \, {\rm MeV}$.}
\label{fig:4}
\end{center}
\end{figure}

In Figs.~\ref{fig:2}, \ref{fig:3}, and \ref{fig:4} we present our results for sterile neutrinos produced via oscillations with electron, muon, and tau neutrinos, respectively, taking ${\rm sin}^2 2\theta =0.1$. We show $n_{\nu_s}/n_{\nu_\alpha}$ as a function of sterile neutrino mass $m_{\nu_s}$ for the standard cosmology (blue), the kination model with $T_{tr}=4\,{\rm MeV}$ (red), the scalar-tensor model (green), and the low reheating temperature model with $T_R=4\,{\rm MeV}$ (purple).   The solid lines show the number density found by integrating Eq.~\ref{extended}, while the dashed lines show the approximate analytic solutions assuming $f_{\nu_s}\ll f_{\nu_\alpha}$.  This approximation ignores the back reaction $\nu_s \rightarrow \nu_{\alpha}$, and breaks down as  as $f_{\nu_s}$ approaches $f_{\nu_\alpha}$.  The kink in the lines at $m_{\nu_s}\simeq 3\,{\rm keV}$ is due to approximating the QCD phase transition by an instantaneous transition in which $g_\star=90$ goes to $g_\star=10.75$.  The new kink at $m_{\nu_s}\simeq 3 \times 10^{-3}\,{\rm keV}$ ($m_{\nu_s}\simeq 2\,{\rm keV}$) in Fig.~\ref{fig:3} (Fig.~\ref{fig:4}) is due to the transition from $B=10.88\times 10^{-9}$ to $B=3.02\times 10^{-9}$ at $T\simeq 20 \, {\rm MeV}$ ($T\simeq 180 \, {\rm MeV}$), when $\alpha = \mu$ ($\alpha=\tau$) decouples from the thermal bath, again approximated as instantaneous.

Since the number density of sterile neutrinos far from equilibrium is proportional to ${\rm sin}^2 2\theta$, changing the value of the mixing angle simply raises or lowers the dashed lines shown in Figs.~\ref{fig:2}, \ref{fig:3}, and \ref{fig:4}.  For example, decreasing ${\rm sin}^2 2\theta$ to $0.01$ lowers the position of these lines in Figs.~\ref{fig:2}, \ref{fig:3}, and \ref{fig:4} by a factor of 10.  For any value of ${\rm sin}^22\theta$ the solid lines, which are valid when $f_{\nu_s}\simeq f_{\nu_\alpha}$, follow the dashed lines until they approach thermal equilibrium with the active neutrinos, at which point they flatten out to $n_{\nu_s}/n_{\nu_\alpha}\simeq1$. 

\section{Cosmological limits on sterile neutrino masses and mixings}

\begin{figure}
\begin{center}
\includegraphics[width=9.5cm]{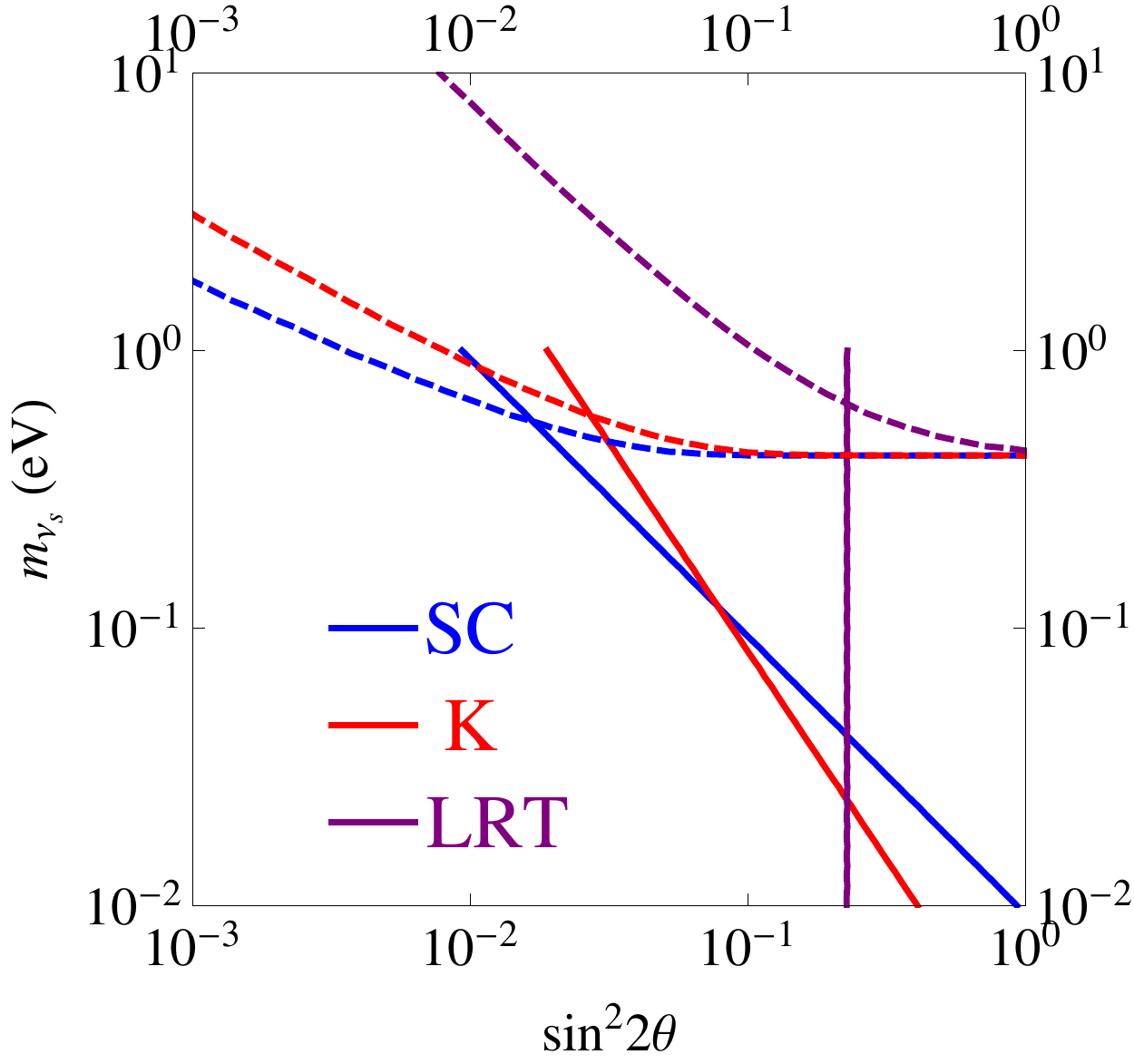}
\caption{Upper limits on the sterile neutrino mass $m_{\nu_s}$ as a function of the active-sterile mixing ${\rm sin}^22\theta$ imposed by the limits on $N_{eff}$ in Eq.~\ref{nval} (solid lines) and $m_{\nu_s}^{eff}$ in Eq.~\ref{mval} (dashed lines) for the standard cosmology (blue line), the kination model with $T_{tr}=4\,{\rm MeV}$, and the low reheating temperature model with $T_R=4\,{\rm MeV}$ for sterile neutrinos mixing with electron neutrinos.  The entire region shown in the plot is allowed in the scalar-tensor model with $T_{tr}=4\,{\rm MeV}$.}
\label{ebounds}
\end{center}
\end{figure}

\begin{figure}
\begin{center}
\includegraphics[width=9.5cm]{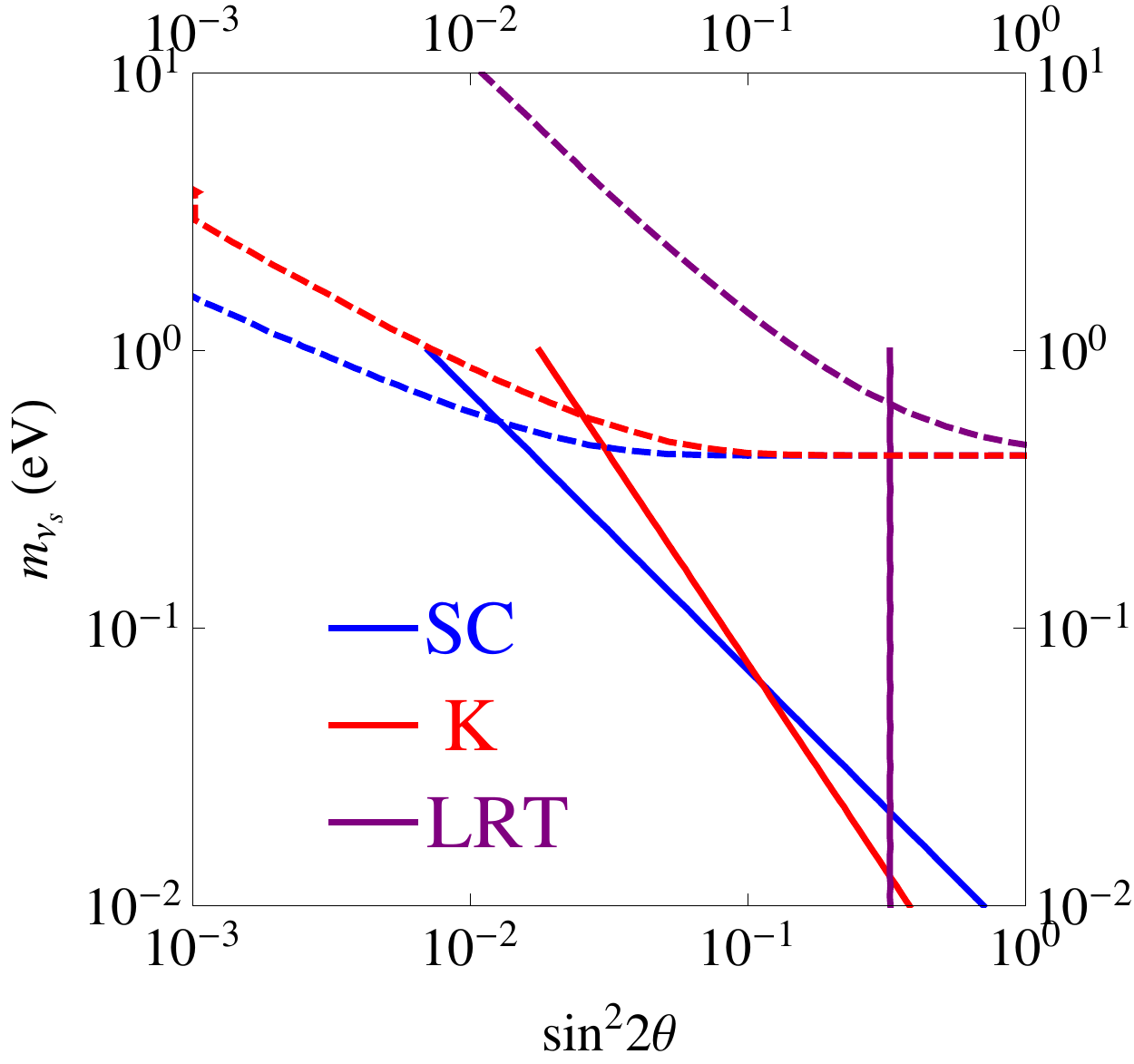}
\caption{Same as Fig.~\ref{ebounds}, but for sterile neutrinos mixing with muon neutrinos.}
\label{ubounds}
\end{center}
\end{figure}

\begin{figure}
\begin{center}
\includegraphics[width=9.5cm]{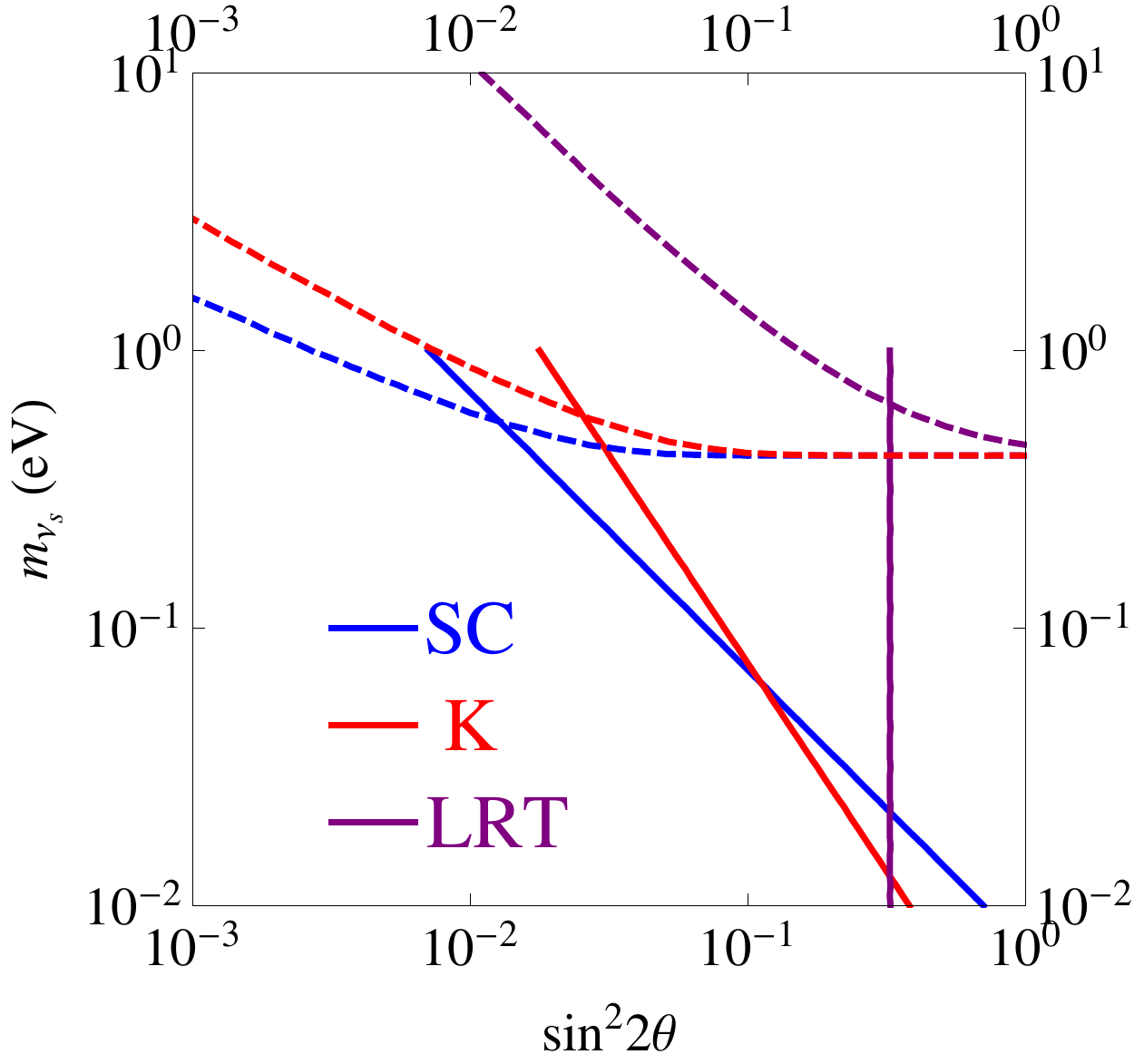}
\caption{Same as Fig.~\ref{ebounds}, but for sterile neutrinos mixing with tau  neutrinos.}
\label{tbounds}
\end{center}
\end{figure}

As can be seen in Figs.~\ref{fig:2}, \ref{fig:3}, and \ref{fig:4}, the relic number density of sterile neutrinos produced in non-standard pre-BBN cosmologies can be significantly smaller than the number produced in the standard cosmology.  As shown in Figs.~\ref{ebounds}, \ref{ubounds}, and \ref{tbounds}, this leads to a relaxation of the bounds on the mixing angle derived from the upper limits on $N_{eff}$ and $m_{\nu_s}^{eff}$ given in Eqs.~\ref{nval} and \ref{mval}, respectively.  We consider sterile neutrinos with mass larger than $0.01 \,{\rm eV}$ to ensure that they are produced during the non-standard pre-BBN cosmological phase.  The bound on $N_{eff}$ applies only to sterile neutrinos that are relativistic at the time of recombination, thus we numerically integrated Eq.~\ref{extended} to find $\rho_{\nu_s}$ in Eq.~\ref{neff}, assuming that sterile neutrinos of mass $m_{\nu_s}<1\,{\rm eV}$ are entirely relativistic. The bound on $m_{\nu_s}^{eff}$ applies to sterile neutrinos which are non-relativistic at recombination, thus their energy density is $\rho_{\nu_s}=m_{\nu_s}n_{\nu_s}$.  We again numerically integrated Eq.~\ref{extended} to compute $n_{\nu_s}$ in Eq.~\ref{meff}.  Applying the limits in Eqs.~\ref{nval} and \ref{mval} to neutrinos with $m_{\nu_s}<10\,{\rm eV}$, we obtain the upper limits presented in Figs.~\ref{ebounds}, \ref{ubounds}, and \ref{tbounds}.

Figs.~\ref{ebounds}, \ref{ubounds}, and \ref{tbounds} show the upper limits on the active sterile mixing ${\rm sin}^22\theta$ as a function of the sterile neutrino mass $m_{\nu_s}$ for the standard cosmology (blue line), kination model with $T_{tr}=4\,{\rm MeV}$, and the low reheating temperature model with $T_R=4\,{\rm MeV}$, for sterile neutrinos mixing with electron, muon, and tau neutrinos, respectively.  The dashed lines come from the bound on $m_{\nu_s}^{eff}=n_{\nu_s}m_{\nu_s}/n_{\nu_\alpha}$ in Eq.~\ref{mval}, while the solid lines are from the bound on $N_{eff}$ in Eq.~\ref{nval}.  The entire region of parameter space in the figures is allowed in the scalar-tensor model with $T_{tr}=4\,{\rm MeV}$, i.e. both limits in Eqs.~\ref{nval} and \ref{mval} are fulfilled for all values of the mixing and $m_{\nu_s}<10\,{\rm eV}$.

Our results show that the LSND $\nu_s$, i.e. a neutrino of mass $m_{\nu_s}\simeq 1 \,{\rm eV}$ and mixing angle ${\rm sin}^22\theta\simeq 0.08$, is allowed by the bounds in Eqs.~\ref{nval} and \ref{mval} in the scalar-tensor and low reheating temperature pre-BBN cosmologies, while such a sterile neutrino would be disfavored in the standard and kination cosmologies.  Therefore, if this neutrino does exist, its discovery could be a sign that the Universe evolves differently than in the standard cosmology prior to BBN, either in a low reheating temperature or a scalar-tensor pre-BBN cosmology.

\begin{figure}
\begin{center}
\includegraphics[width=9.5cm]{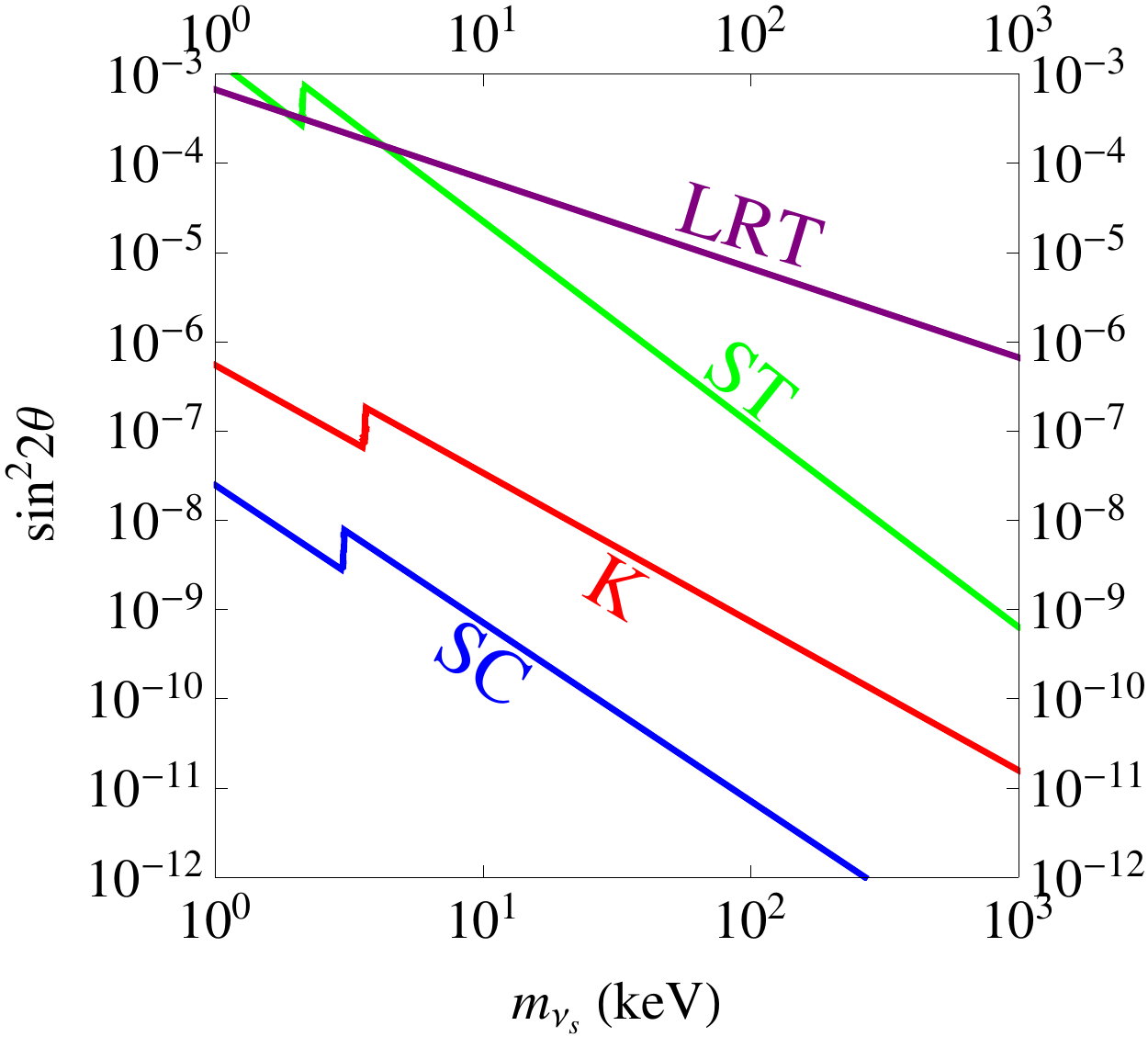}
\caption{Upper limits on the sterile-active neutrino mixing ${\rm sin}^22\theta$ as a function of the sterile neutrino mass $m_{\nu_s}$ for $m_{\nu_s}\gtrsim {\rm keV}$ in the standard cosmology (blue line),the kination (red line) and scalar-tensor (green line) models with $T_{tr}=4\,{\rm MeV}$, and the low reheating temperature model with $T_R=4\,{\rm MeV}$ for sterile neutrinos mixing with electron neutrinos.}
\label{kevebounds}
\end{center}
\end{figure}

Sterile neutrinos of mass $m_{\nu_s}\gtrsim {\rm keV}$, that could make up all (or some large fraction) of the dark matter, must have a relic energy density less than the energy density of dark matter, i.e. $\Omega_{\nu_s}=\rho_{\nu_s}/\rho_{crit}\leq \Omega_{DM}$, where $\rho_{crit}$ is the critical density of the Universe.  To compute this bound, we use $\Omega_{DM} h^2=0.1198$, $\rho_{crit}=1.0538 h^2\,{\rm GeV/cm}^3$, and $h=0.673$. \cite{Beringer}.

\begin{figure}
\begin{center}
\includegraphics[width=9.5cm]{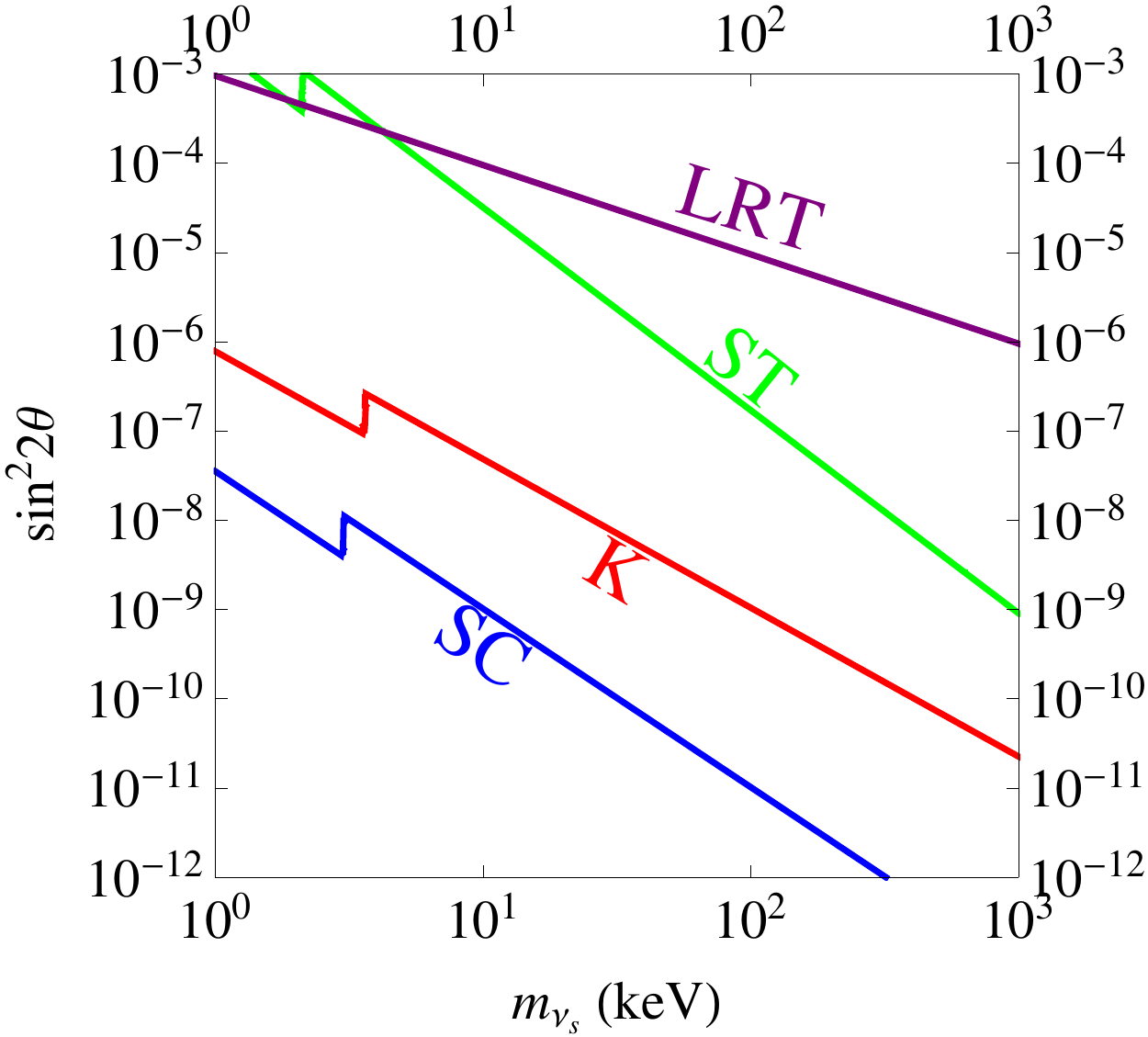}
\caption{Same as Fig.~\ref{kevebounds}, but for sterile neutrinos mixing with muon neutrinos.}
\label{kevubounds}
\end{center}
\end{figure}

\begin{figure}
\begin{center}
\includegraphics[width=9.5cm]{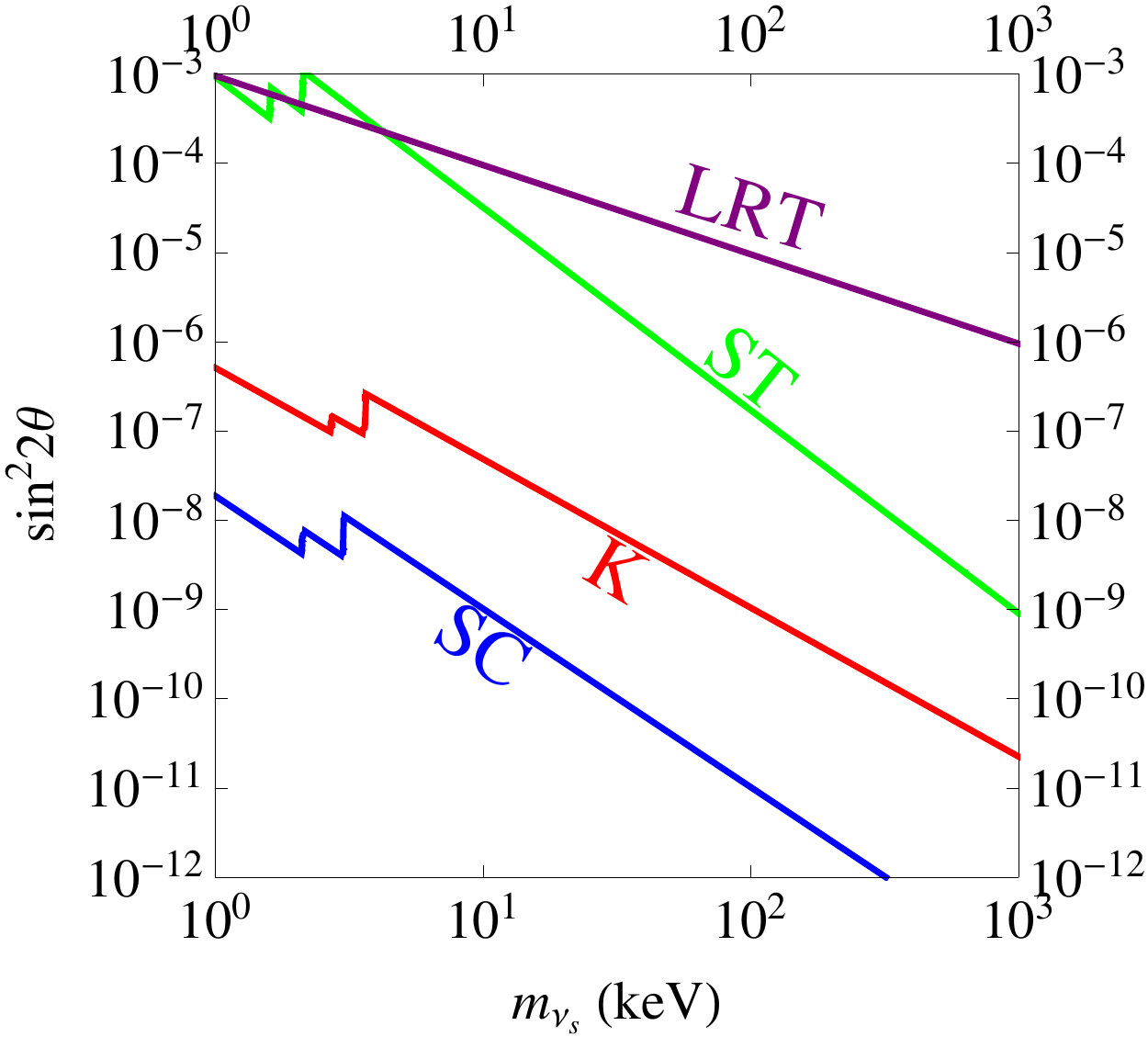}
\caption{Same as Fig.~\ref{kevebounds}, but for sterile neutrinos mixing with tau neutrinos.}
\label{kevtbounds}
\end{center}
\end{figure}

Figs.~\ref{kevebounds}, \ref{kevubounds}, and \ref{kevtbounds} show the upper limits on ${\rm sin}^22\theta$ as a function of $m_{\nu_s}$ due to $\Omega_{\nu_s}<\Omega_{DM}$ for sterile neutrinos with $m_{\nu_s}\gtrsim {\rm keV}$ in the standard cosmology (blue line), the kination (red line) and scalar-tensor (green line) models with $T_{tr}=4\,{\rm MeV}$, and the low reheating temperature model with $T_R=4\,{\rm MeV}$, for sterile neutrinos mixing with electron, muon, and tau neutrinos, respectively.  As can be seen in Figs.~\ref{kevebounds}, \ref{kevubounds}, and \ref{kevtbounds}, the upper limit on ${\rm sin}^22\theta$ for each mass $m_{\nu_s}$ is relaxed in the non-standard pre-BBN cosmologies we consider since there are fewer sterile neutrinos produced than in the standard cosmology.  Since X-ray measurements, phase space density arguments and Lyman-$\alpha$ forest measurements apply to sterile neutrinos making up most of the dark matter, non-standard pre-BBN cosmologies open a new region in parameter space for sub-dominant sterile neutrinos.  The experimental discovery of a sterile neutrino forbidden in the standard cosmology and allowed in others could, therefore, be evidence of a non-standard pre-BBN cosmology.

\section{General constraints on H}

Here we present some general constraints on the functional dependence of H on the temperature that guarantee the suppression of the present number density of sterile neutrinos produced via the Dodelson-Widrow mechanism.  We consider non-standard expansion rates with $H_{NS}\sim T^n$ for $n \geq 0.$  If $H_{NS}$ changes smoothly into the standard expansion rate at $T_{tr}$, then $H_{NS}\simeq \pi T^n/(M_P T_{tr}^{n-2})\sqrt{g_\star / 90}.$  In order for sterile neutrino production to remain out of equilibrium, the interaction rate of sterile neutrinos (given in Eq.~\ref{rate}) must always be below the expansion rate of the Universe, i.e. $\Gamma_s<H.$
To ensure that this is satisfied, the peak interaction rate must be
\begin{equation}
\Gamma_s^{peak}(T_{max})<H_{NS}(T_{max}),
\label{cond}
\end{equation}
where $T_{max}$ is the temperature that the peak occurs at (see Sect.~\ref{3}).  This condition can be rewritten to find a lower bound on $n,$
\begin{equation}
n>{\rm ln}\left(\frac{1}{\pi}\left(\frac{90}{g_\star}\right)^{1/2}\frac{\Gamma_s^{peak}M_P}{T_{tr}^2}\right)\left[{\rm ln}\left(\frac{T_{max}}{T_{tr}}\right)\right]^{-1}.
\end{equation}
For an LSND $\nu_s$ coupled to electron neutrinos, and taking the minimum possible value for the transition temperature, $T_{tr}=4\,{\rm MeV},$ we find that $n>4.7.$  This justifies that, as we have proven before, an LSND $\nu_s$ enters into equilibrium in a kination cosmology since $n=3.$  In addition, we note that in order to suppress the production of sterile neutrinos the transition temperature must be smaller than the temperature at which the interaction rate drops below the standard cosmology expansion rate, i.e. $T_{tr}<T_{SC}$, where $T_{SC}$ is defined by $H_{SC}(T_{SC})=\Gamma_s(T_{SC})$.  For an LSND $\nu_s$ this is $T_{SC}=5.3\,{\rm MeV}$ (see Fig.~\ref{fig:Hubble}).  If $T_{tr}=T_{SC},$ Eq.~\ref{cond} requires $n>5.$

We have so far considered a smooth transition between $H_{NS}$ and $H_{SC}.$  If $H_{NS}$ has a quick jump at $T_{tr}$, as in the scalar-tensor cosmology, we can find an approximate condition on $H_{NS}$ for small $n$ by considering $H_{NS}$ as constant during the period in which the sterile neutrino production rate is maximum, $H_{NS}=H_{NS}(T_{max}).$  Under this approximation, we can solve Eq.~\ref{explain1}, and find that the ratio $n_{\nu_s}/n_{\nu_\alpha}<r,$ where r is a number between $0$ and $1,$ when
\begin{equation}
\frac{H_{NS}(T_{max})}{H_{SC}^{m_{\nu_s}=1\,{\rm eV}}(T_{max})}>98\, d_\alpha {\rm sin}^2(2\theta)\left( \frac{m_{\nu_s}}{{\rm eV}}\right)^{5/3}\left(\frac{10.88\times 10^{-9}}{B}\right)^{5/6}r^{-1},
\end{equation}
where $H_{SC}^{m_{\nu_s}=1\,{\rm eV}}(T_{max})=8.74\times 10^{-23}\,{\rm GeV}$ is the value of the standard cosmology expansion rate at $T_{max}$ for sterile neutrinos with $m_{\nu_s}=1\,{\rm eV}$.  For the production of the LSND $\nu_s$ coupled to electron neutrinos to never be in equilibrium, $r=1$, so $H_{NS}(T_{max})>8.9\,H_{SC}^{m_{\nu_s}=1\,{\rm eV}}(T_{max})$.  In order for the LSND $\nu_s$ to be allowed by the Planck joint bounds in Eqs.~\ref{nval} and \ref{mval} (for $m_{\nu_s} = 1\,{\rm eV}$, Eq.~\ref{mval} is more constraining), $r=0.42$, leading to $H_{NS}(T_{max})>21\,H_{SC}^{m_{\nu_s}=1\,{\rm eV}}(T_{max}).$

\section{Conclusions}

In this paper we presented the effects of some non-standard pre-BBN cosmologies on the production of sterile neutrinos.  If the peak of the production rate of sterile neutrinos occurs during the non-standard cosmological phase, we have shown that the number density of sterile neutrinos produced non-resonantly could be much smaller than in the standard cosmology. Thus, current bounds on the active-sterile neutrino mixing derived from cosmological bounds on the energy density of sterile neutrinos can be greatly relaxed if the Universe developed differently than the standard radiation dominated cosmology assumes.  We find that if the Universe evolves in a scalar-tensor or low reheating temperature cosmology prior to BBN, a sterile neutrino that could explain the anomalies observed in short-baseline neutrino experiments would be allowed by the Planck joint bounds in Eq.~\ref{nval} and \ref{mval}, although it would be disfavored if the Universe evolved according to the standard or kination cosmologies.  In general, if the expansion rate during the non-standard cosmological phase depends on temperature as $T^n$, with $n\geq 0,$ and goes smoothly into the standard cosmology, the LSND $\nu_s$ production would never be in equilibrium only if $n>4.7.$  If the non-standard expansion rate has a quick jump at the transition into the standard cosmology, and has a very small $n,$ the condition is instead that the expansion rate in the non-standard cosmology at the time of the peak interaction rate must be $8.9$ times that of the standard cosmology, and to evade the Planck joint bounds, this number should be $21.$

We also find that the kination, scalar-tensor, and low reheating temperature cosmologies open up a new region of phase space for sterile neutrinos of mass $m_{\nu_s}\gtrsim {\rm keV}$ that do not make up all of the dark matter.  The experimental discovery of such a neutrino could, therefore, be evidence of a non-standard pre-BBN cosmology.

\section*{Acknowledgements}

G.G. was supported in part by the  Department of Energy under Award Number DE-SC0009937.

\end{document}